\documentclass{iopart}

\usepackage{iopams}

\usepackage{multicol}
\usepackage{graphicx}

\begin{document}

\title[High-temperature signatures of quantum criticality]{High-temperature signatures of 
quantum criticality in heavy-fermion systems}

\author{J. Kroha$^1$, M. Klein$^2$, A. Nuber$^2$, F. Reinert$^{2,3}$, \\
O. Stockert$^4$ and H. v. L\"ohneysen$^{5,6}$
}

\address{ $^1$
Physikalisches Institut and Bethe Center for Theoretical Physics,\\
Universit\"at Bonn, Nussallee 12, 53115 Bonn, Germany
}\vspace*{-1ex}\ead{kroha@physik.uni-bonn.de}

\address{ $^2$
Universit\"at W\"urzburg, Experimentelle Physik II,  Am Hubland,\\ 
97074 W\"urzburg, Germany
}

\address{ $^3$
Forschungszentrum Karlsruhe, Gemeinschaftslabor f\"ur Nanoanalytik,\\ 
76021 Karlsruhe, Germany
}

\address{ $^4$
Max Planck Institute for Chemical Physics of Solids, N\"othnitzer Str. 40,\\
01187 Dresden, Germany
}

\address{ $^5$
Physikalisches Institut, Universit\"at Karlsruhe, 76128 Karlsruhe, Germany
}

\address{ $^6$
Forschungszentrum Karlsruhe, Institut f\"ur Festk\"orperphysik, 
76021 Karlsruhe, Germany
}

\begin{abstract}
We propose a new criterion for distinguishing the Hertz-Millis (HM)
and the local quantum critical (LQC) mechanism in heavy-fermion systems
with a magnetic quantum phase transition (QPT). The criterion is based on
our finding that the complete spin screening of Kondo ions can be 
suppressed by the RKKY coupling to the surrounding magnetic ions even without 
magnetic ordering and that, consequently, the signature of this suppression
can be observed in spectroscopic measurements above the magnetic 
ordering temperature. We apply the criterion to high-resolution 
photoemission measurements on CeCu$_{6-x}$Au$_{x}$ and conclude that 
the QPT in this system is dominated by the LQC scenario. 
\end{abstract}


\section{Introduction}
\label{intro}
\begin{figure}[t]
\begin{center} \scalebox{0.25}[0.25]{\includegraphics[clip]{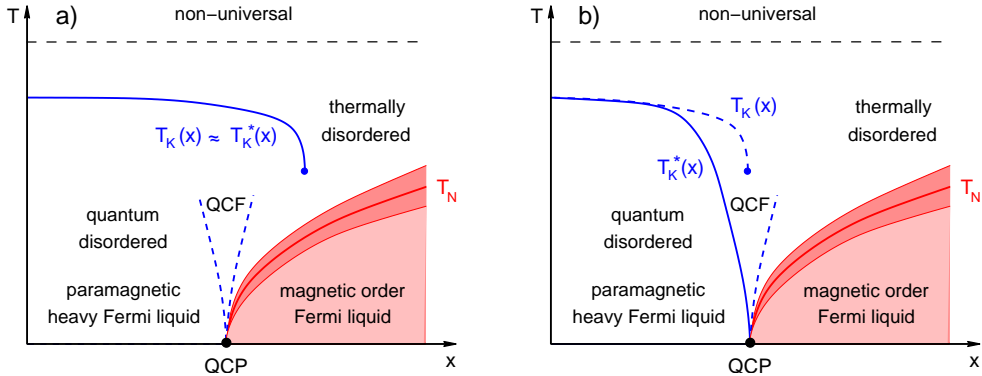}}\end{center}
\vspace*{-0.3cm}
\caption{
Generic phase diagrams of a magnetic QPT in a HF system,
driven by an antiferromagnetic RKKY coupling (parametrized by a non-thermal 
control parameter $x$) 
for a) the HM and b) the LQC scenario. For both scenarios
the predicted behavior of the spin screening scale on the lattice, 
$T_K^{\star}$, {\it including} the presence of 
quantum critical fluctuations (QCF), 
and as extracted from local Kondo-ion
spectra {\it without} lattice coherence or QCF, $T_K$, is also shown 
(see text for details).
The maximum antiferromagnetic RKKY coupling, $x_m$, 
where single-ion Kondo screening terminates, is marked by a black dot. 
}
\label{fig1}
\end{figure}

At a second-order phase transition the characteristic time scale of the
order parameter fluctuations diverges (critical slowing down), because 
the energy difference between the ordered and the disordered phases, i.e.,
the fluctuation energy $\omega_{fl}$, vanishes continuously at the 
transition. If the phase transition occurs at a finite critical 
temperature $T_c$, quantum fluctuations of the order-parameter are always 
cut off by the temperature $T$, since $T\approx T_c > \omega_{fl}$, and the
order-parameter fluctuations are thermally excited, i.e., incoherent 
(dark shaded regions in Fig.~\ref{fig1} a), b)). 
In this sense, such a phase transition is classical. 
If, however, the transition is tuned to absolute zero temperature by a 
non-thermal control parameter, the system is at the critical point 
in a quantum coherent 
superposition of the degenerate ordered and disordered states.
The transition is called a quantum phase transition (QPT),
for reviews see \cite{loehneysen07,steglich08}.  
The excitation spectrum above this quantum critical state may be distinctly 
different from the excitations of either phase, the disordered and the 
ordered one. Therefore, the physical properties are not only dominated 
by the quantum fluctuations between these phases at $T=0$, but show also
unusual temperature dependence essentially due to thermal excitation of the 
anomalous spectrum, so that the quantum critical behavior extends up to
elevated temperatures (regions marked "QCF" in Fig.~\ref{fig1} a), b)).

%%%%%%%%%%%%%%%%%%%%%%%%%%%%%%%%%%%%%%%%%%%%%%%%%%%%%
%%%%%%%%%%%%%%%%%%%%%%%%%%%%%%%%%%%%%%%%%%%%%%%%%%%%%

In particular, in a number of heavy-fermion (HF) compounds, where heavy
quasiparticles are formed due to the
Kondo effect and subsequent lattice coherence, a magnetic phase 
transition may be suppressed to $T=0$ by chemical composition, pressure
or magnetic field. Two types of scenarios are in principle conceivable in these
metallic systems.\\
In the first scenario, the quasiparticle system  
undergoes a spin-density wave (SDW) instability at the quantum critical 
point (QCP), as decribed by the theory of 
Hertz \cite{hertz76} and Millis \cite{millis93} (HM scenario). 
The instability can be caused by various types of residual spin exchange 
interactions. In this scenario the Landau Fermi liquid, albeit 
undergoing magnetic 
ordering, prevails, and the Kondo temperature $T_K$ remains finite across 
the QPT. \\
In the second type of scenario, the Kondo effect and, hence, the very formation
of heavy fermionic quasiparticles is suppressed. This may occur due 
to magnetic coupling to the surrounding moments \cite{si01,coleman01}
or possibly due to fluctuations of the Fermi volume involved with the 
onset of Kondo screening in an Anderson lattice system \cite{senthil04}. 
Both the bosonic order-parameter fluctuations and the 
local fermionic excitations then become critical at the 
QPT \cite{si01,coleman01}. 
In this case the system is in a more exotic, genuine many-body state which is 
not described by the Landau Fermi-liquid paradigm. For the critical 
breakdown of Kondo screening due to magnetic fluctuations 
the term ``local quantum critical (LQC)'' has been coined \cite{si01}.

Unambiguously identifying the quantum critical scenario from the low-$T$
behavior, not to speak of predicting the scenario for a given system, 
has remained difficult. One reason for this is that the precise critical
behavior is not known because of approximate assumptions implicit in the
theoretical description of either one scenario, HM or LQC.
While the HM theory \cite{hertz76,millis93} pre-assumes the existence of
fermionic quasiparticles with only bosonic, critical order-parameter 
fluctuations, the extended dynamical mean field theory (EDMFT) description
of the LQC scenario \cite{si01,coleman01} 
neglects possible changes of the critical behavior due 
to spatially extended critical fluctuations. 
Motivated by our recent ultraviolet 
(UPS) \cite{klein08} and X-ray (XPS) \cite{klein09}
photoemission spectroscopy measurements of the Kondo resonance 
at elevated $T$ across the Au-concentration range of the QPT 
in CeCu$_{6-x}$Au$_x$, we here put forward a criterion 
to predict the quantum critical scenario of a HF system from its 
high-$T$ behavior around and above the single-ion Kondo temperature $T_K$. 
As seen below, this criterion derives from the fact that the complete 
Kondo screening breaks down when the dimensionless RKKY coupling $y$ 
between Kondo ions exceeds a certain strength $y_{m}$, even when critical 
fluctuations due to magnetic ordering do not play a role.
$y_m$ can be expressed in a universal way in terms of the bare 
single-ion Kondo temperature $T_K(0)$ only, see Eq.(\ref{eq:ymax}) below.  
This breakdown is related to the 
unstable fixed point of the two-impurity Kondo model which separates the 
Kondo-screened and the inter-impurity (molecular) singlet ground states of 
this model \cite{jones87}. 
In the present paper we explore and utilize its signatures at 
temperatures well above the lattice coherence temperature $T_{coh}$ 
and the magnetic-ordering or N\'eel temperature $T_N$, i.e., in a
region where neither critical fluctuations of the Fermi 
surface \cite{senthil04} or the magnetic order parameter play a role.  

In the following Section \ref{theory} we present our calculations of the 
high-temperature signatures of the RKKY-induced Kondo breakdown using 
perturbative renormalization group as well as selfconsistent diagrammatic 
methods. In Section \ref{experiment} we briefly recollect the UPS results for 
CeCu$_{6-x}$Au$_x$ \cite{klein08}
and interpret them in terms of the high-$T$ signatures of Kondo breakdown.
Some general conclusions are drawn in Section \ref{conclusion}.

\section{Theory for single-ion Kondo screening in a Kondo lattice}
\label{theory}
\begin{figure}[t]
\begin{center} \scalebox{0.305}[0.305]{\includegraphics[clip]{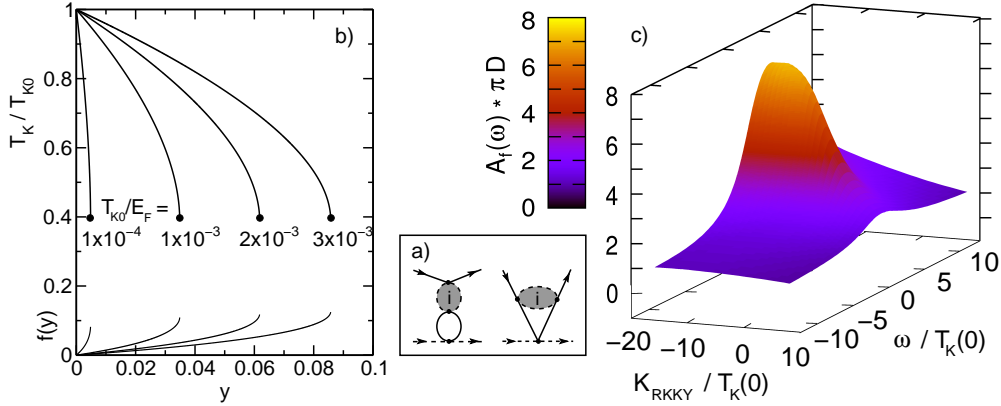}}\end{center}
\vspace*{-0.3cm}
\caption{
a) Leading order RKKY-induced corrections to the local spin-exchange coupling.
Solid and dashed lines represent conduction electron and impurity spin
(pseudofermion) propagators, respectively. 
b) The single-impurity Kondo scale $T_K(y)$ with RKKY 
corrections to the local exchange coupling (Fig.~\ref{fig2} a))
is shown as a function of $y$
for various values of the bare Kondo temperature $T_K(0)$.
The effective perturbation parameter $f(y)$ is also shown.
c) NCA result for the f spectral density on a single Kondo ion, 
including RKKY corrections (Fig.~\ref{fig2} a)) for $T=2\, T_K(0)$. 
The steep collapse of the Kondo resonance for increasing antiferromagnetic
RKKY coupling $K_{RKKY}$ is clearly seen.
}
\label{fig2}
\end{figure}

We consider a HF system described by the Kondo lattice model 
of local 4f spins $S=1/2$
with the exchange coupling $J$ to the conduction electrons 
and the density of states at the
Fermi level, $N(0)$, 
for temperatures well above  $T_{coh}$, $T_N$. In this regime
controlled calculations of renormalized perturbation theory in terms of
the single-impurity Kondo model are possible and 
can be directly compared to experiments \cite{klein08}. In particular, the
RKKY interaction of a given Kondo spin at site 0 with identical spins at the
surrounding sites $i$ can be treated as a perturbative correction to the local 
coupling $J$. The leading-order direct and exchange 
corrections $\delta J^{(d)}$, $\delta J^{(ex)}$ are depicted diagrammatically 
in Fig.~\ref{fig2} a). As seen from the figure, these corrections involve the 
full dynamical impurity spin susceptibility (shaded bubbles) on the 
neighboring impurity sites $i$,
$\chi_{4f}(T,0)\!=\!(g_L\mu_B)^2\,N(0) D_0/(4\sqrt{T_K^{2}+T^2})$,
with the bare band width $D_0\approx E_F$ and
the Land\'e factor $g_L$ and the Bohr magneton $\mu_B$
\cite{andrei83}. Note that the Kondo temperature
of this effective single-impurity problem, $T_K$, is to be 
distinguished from the spin screening scale of the lattice problem
\cite{pruschke00}, $T_K^{\star}$, which would also include QCF. 
Summing over all lattice sites $i\neq 0$ one obtains \cite{klein08},
\begin{eqnarray}
\delta J^{(d)} &=& - y \frac{1}{4} J g_{i}^2 \ \frac{D_0}{\sqrt{T_K^{2}+T^2}}\
\frac{1}{1+(D/T_K)^2}
\label{eq:deltaJ_d}\\
\delta J^{(ex)} &=& - y \frac{1}{4} J g_{i}^2 
\left( \frac{3}{4} + \frac{T}{\sqrt{T_K^{2} + T^2}} \right)\ .
\label{eq:deltaJ_ex}
\end{eqnarray}
Here $g_i\!=\!N(0)J_i$ is the dimensionless bare coupling on site $i\!\neq\!0$,
$y$ is a dimensionless factor that describes the 
relation between the RKKY coupling strength and the Au content $x$. 
In the vicinity of the QPT we assume a linear dependence, $y=\alpha (x+x_0)$,
with adjustable parameters $\alpha$ and $x_0$.
In $\delta J^{(d)}$ [first diagram in Fig.~\ref{fig1} a)] the local spin 
response $\chi_{4f}(T,\Omega)$ 
restricts the energy exchange between conduction electrons and local spin,
i.e. the band cutoff $D$, to $T_K$. This is described by the last 
factor in Eq.~(\ref{eq:deltaJ_d}) (soft cutoff).
In $\delta J^{(ex)}$ [second diagram in Fig.~\ref{fig1} a)]  
$\chi_{4f}(T,\Omega)$ restricts the conduction-electron
response to a shell of width $T_K$ around the Fermi energy $E_F$, 
and suppresses $\delta J^{(ex)}$ compared to
$\delta J^{(d)}$ by an overall factor of $\sqrt{T_K^{2}+T^2}/D_0$, 
as seen in Eq.~(\ref{eq:deltaJ_ex}). 
The spin screening scale of this effective 
single-impurity problem, including RKKY corrections, can now be 
obtained as the energy scale where the perturbative renormalization group
(RG) for the RKKY-corrected spin coupling (taken at $T=0$) diverges. 
The one-loop RG equation reads
\begin{eqnarray} 
\frac{d J}{d\ln D}\!=\!-2 N(0) 
\left[ J+\delta J^{(d)}(D)+\delta J^{(ex)}(D) \right]^2 \ .
\label{eq:RGequation}
\end{eqnarray}
Note that in this RG equation the bare bandwidth $D_0$ and the 
couplings $g_i$ on sites $i\!\neq\!0$ are not renormalized, 
since this is already included in the full susceptibility $\chi_{4f}$. 
The essential feature is that for $T=0$ the direct RKKY correction
$\delta J^{(d)}$, Eq.~(\ref{eq:deltaJ_d}), is inversely proportional to the 
renormalized Kondo scale $T_K(y)$ itself via $\chi_{4f}(0,0)$.
The solution of Eq.~(\ref{eq:RGequation})
leads to a highly non-linear self-consistency equation for $T_K(y)$,
%%%%%%%%%%%%%%%%%%%%%%%%%%%%%%%%%%%%%%%%%%%%%%%%%%%%%
%%%%%%%%%%%%%%%%%%%%%%%%%%%%%%%%%%%%%%%%%%%%%%%%%%%%%
\begin{eqnarray}
\frac{T_K(y)}{T_K(0)} = \exp
\left\{-\left(\frac{1}{2g} +\ln 2\right) \
\frac{f(u)}{1-f(u)}
\right\}  \ ,
\label{eq:TK_RG}
\end{eqnarray}
with $g\!=\!N(0)J$, $f(u)\!=\!u-u^2/2$, $u\!=\!yg^2D_0/[4T_K(y)]$.
The single-ion Kondo scale without RKKY coupling is
$T_K(0)=D_0\ {\rm exp}[-1/2g]$.
Fig.~\ref{fig2} b) shows solutions of Eq.~(\ref{eq:TK_RG}) for various 
values of $T_K(0)$, together with the corresponding values of the effective 
perturbation parameter $f(y)$. It is seen that this RG treatment
is perturbatively controlled in the sense that $f(y)\lesssim 0.1$, i.e., 
the exponent in Eq.~(\ref{eq:TK_RG}) remains small for all solutions. 
Remarkably, a solution of Eq.~(\ref{eq:TK_RG}) exists only up to a certain
RKKY coupling strength $y_m$. For each value of the bare single-ion 
Kondo temperature, $\tau_K=T_K(0)/D_0$,  $y_m$ can be calculated
as the point where the derivative $dT_K(y)/dy$ diverges \cite{klein08}, 
\begin{eqnarray}
y_{m}=3.128\,\tau_K\,(\ln \tau_K)^2\hspace{-0.2ex}
\left[2\hspace{-0.2ex}-\hspace{-0.2ex}\ln\frac{\tau_K}{2}\hspace{-0.2ex}-
\hspace{-0.2ex}\sqrt{\left(2\hspace{-0.2ex}-\hspace{-0.2ex}
\ln\frac{\tau_K}{2}\right)^2\hspace{-0.2ex}-4} 
\right] \ .
\label{eq:ymax}
\end{eqnarray}
By rescaling $y$ and $T_K(y)$ as $y/y_m$ and
$T_K(y)/T_K(0)$, respectively, all $T_K(y)$ curves 
collapse onto a single, universal curve, shown in the inset 
of Fig.~\ref{fig5} below. 
For $y>y_m$ the RG Eq.~(\ref{eq:RGequation}) does not diverge, 
i.e., the Kondo screening breaks down at this maximum RKKY coupling 
strength $y_m$, even if magnetic ordering does not occur. Therefore, the
physical origin of the high-$T$ criterion (\ref{eq:ymax}) is different 
from the well-known Doniach criterion \cite{doniach77} (which reads
$T_K(0)\approx y_m N(0)J^2$),
albeit it yields numerically similar values for $y_m$. 
According to Fig.~\ref{fig2} b) a sharp drop of $T_K$ is predicted 
at $y=y_m$. As seen in Fig.~\ref{fig2} c), this breakdown of Kondo 
screening is signalled by a 
collapse of the Kondo resonance in the local 4f spectrum $A_f(\omega)$ 
of a single Kondo ion, as the antiferromagnetic RKKY coupling to
neighboring Kondo ions is increased.
Fig.~\ref{fig2}~c) shows $A_f(\omega)$ as calculated for the 
two-impurity Anderson model within the non-crossing approximation 
(NCA) at $T=2\,T_K(0)$ \cite{keiter71}. For an efficient 
implementation of the NCA see \cite{costi96}. 
Details of these calculations as well as numerical renormalization group (NRG) 
studies of this problem will be published   
elsewhere. The described signatures should be directly observable 
in spectroscopic experiments at temperatures well above $T_N$, 
see section \ref{experiment}. We emphasize again that the 
Kondo breakdown occurs in any case, whether or not magnetic 
ordering sets in at low $T$. 

%%%%%%%%%%%%%%%%%%%%%%%%%%%%%%%%%%%%%%%%%%%%%%%%%%%%%
\begin{figure}[t]
\begin{center} \scalebox{0.35}[0.35]{\includegraphics[clip]{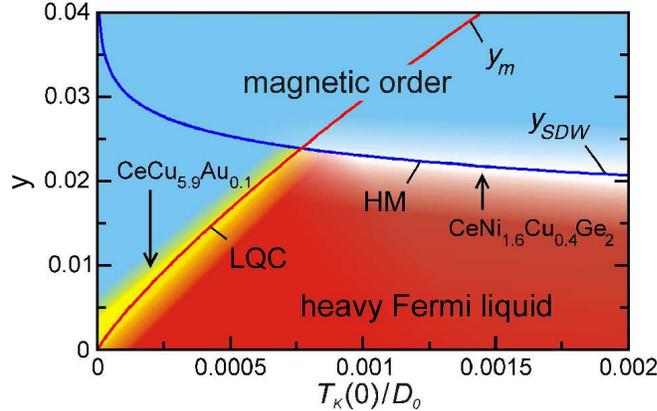}}\end{center}
\vspace*{-0.3cm}
\caption{
Schematic phase diagram of a HF system with a magnetic QPT
in the $T_K(0)$--$y$ plane. The line denoted by $y_m$ represents
Eq.~(\ref{eq:ymax}). At this line $T_K(y)$ undergoes an
abrupt step, see text. The curve denoted by $y_{SDW}$ marks, as an
example, an SDW instability of the system. The magnetic phase transition,
LQC- or HM-like, occurs at whichever of the two lines is lower for a given
system. The arrows indicate estimates \cite{ehm07} for
$T_K(0)/D_0$ for CeCu$_{6-x}$Au$_x$
and CeNi$_{2-x}$Cu$_x$Ge$_{2}$.
  
}
\label{fig3}
\end{figure}
%%%%%%%%%%%%%%%%%%%%%%%%%%%%%%%%%%%%%%%%%%%%%%%%%%%%%

Therefore, the model predicts two quantum critical 
scenarios with distinctly different high-$T$
signatures:  
(1) The heavy Fermi liquid has a magnetic, e.g., SDW instability 
at $T=0$ for an RKKY parameter $y=y_{SDW}<y_m$, i.e., without 
breakdown of Kondo screening. 
In this case, $T_K(y)$, as extracted from high-$T$ UPS spectra,
is essentially constant across the QCP but does have a sharp drop 
at $y=y_m$ inside the region where magnetic ordering occurs at low $T$,
see Fig.~\ref{fig1} a). This corresponds to the HM scenario.
(2) Magnetic ordering does not occur for $y<y_m$. 
In this case the Kondo breakdown at $y=y_m$ implies that 
the residual local moments order at sufficiently low  $T$, i.e., the 
magnetic QCP coincides with $y=y_m$. Quantum critical fluctuations 
(not considered in the present high-$T$ theory) will suppress 
the actual Kondo screening scale $T_K$ below the high-$T$ 
estimate $T_K$, as shown in Fig.~\ref{fig1} b). 
This is the LQC scenario. 
These predictions are summarized in 
Fig.~\ref{fig3} as a phase diagram in terms of the bare 
Kondo scale $T_K(0)$ and the dimensionless RKKY coupling $y$ \cite{klein08}.

%%%%%%%%%%%%%%%%%%%%%%%%%%%%%%%%%%%%%%%%%%%%%%%%%%%%%
\begin{figure}[t]
\begin{center} \scalebox{0.3}[0.3]{\includegraphics[clip]{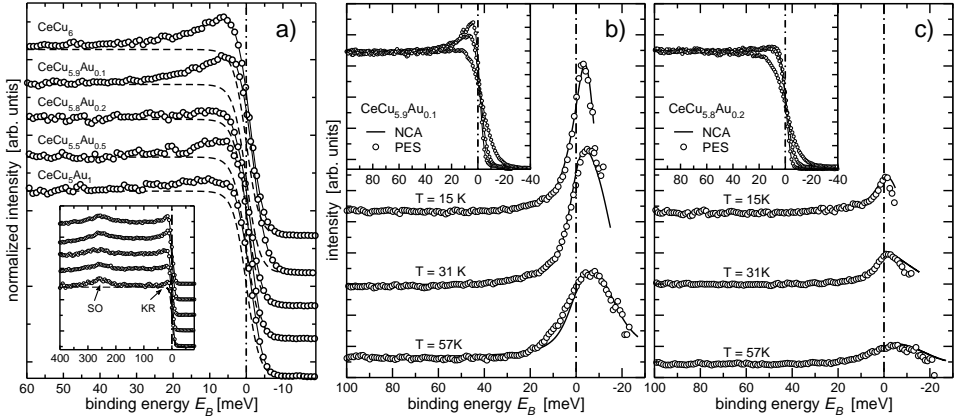}}\end{center}
\vspace*{-0.3cm}
\caption{
{a)} Near-$E_F$ spectra of CeCu$_{6-x}$Au$_x$ for
five different Au concentrations at $T\!=\!15$~K. 
The dashed lines indicate the resolution-broadened
FDD at $T\!=\!15$\,K. The inset shows a larger energy 
range including the spin-orbit (SO) satellite at $E_B\!\approx\!260$~meV.
See Ref.~\cite{klein08} for details of the experimental parameters.   
{b)} and {c)}  show spectra for $x\!=\!0.1$ and
$x\!=\!0.2$, respectively, divided by the
FDD, at various $T$. The solid lines are single-impurity NCA fits.
The insets in {b)} and {c)} show the corresponding raw data.
}
\label{fig4}
\end{figure}
%%%%%%%%%%%%%%%%%%%%%%%%%%%%%%%%%%%%%%%%%%%%%%%%%%%%%

\section{High-resolution photoemission spectroscopy at elevated temperature}
\label{experiment}
The theory described in the previous section should be applicable
quite generally to HF systems with a magnetic QPT. 
Here we apply it to CeCu$_{6-x}$Au$_x$, which is one of the best 
characterized HF compounds
\cite{loehneysen94,loehneysen96a,loehneysen98,stockert98,loehneysen98a,stroka93,stockert07,grube99}.
The Au content $x$ is used to tune the RKKY interaction through 
the QPT at $x=x_c=0.1$, and  
Our recent UPS measurements \cite{klein08} on this compound at 
elevated $T$ have actually motivated the theoretical study.
Details of the sample preparation and measurement procedures can be 
found in \cite{reinert01,ehm07}. The UPS measurements 
were done at $T = 57$\,K, 
31\,K and 15\,K, i.e., well above $T_K(0)\approx 5$\,K, $T_{coh}$, and 
above the temperature up to which quantum critical fluctuations extend in 
CeCu$_{6-x}$Au$_x$ \cite{schroeder00}. Thus they record predominantly the 
local Ce 4f spectral density which is characterized by an 
effective {\it single-ion} Kondo scale $T_K$.
This corresponds to the situation 
for which the calculations in Section \ref{theory} were done.
In Fig.~\ref{fig4} a) raw UPS
Ce 4f spectra are displayed, showing the onset of the Kondo resonance.
A sudden decrease of the Kondo spectral weight at or near the
quantum critical concentration $x_c$ can already be observed in
these raw spectra. The states at energies of up to $~5k_BT$ above the
Fermi level are accessible by a well-established procedure
\cite{reinert01,ehm07} which involves dividing the
raw UPS spectra by the Fermi-Dirac distribution function (FDD). The Kondo 
resonance, which in CeCu$_{6-x}$Au$_x$ is located slightly above the
Fermi level, then becomes clearly visible, see Fig.~\ref{fig4}~b),~c). 
These figures exhibit clearly the collapse of the Kondo spectral 
weight above as compared to below $x_c$. It is in qualitative accordance 
with the Kondo resonance collapse in the theoretical spectra for $T>T_K(0)$, 
Fig~\ref{fig2}~c). To pinpoint the position of the Kondo breakdown more
precisely, the single-ion Kondo temperature $T_K$ was extracted
from the experimental spectra for various $x$. To that end, we followed the 
procedure successfully applied to various Ce compounds 
in the past \cite{patthey90,garnier97,allen00,reinert01,ehm07}:
Using the non-crossing approximation (NCA) 
\cite{costi96} the Ce $4f$ spectral function of the 
single-impurity Anderson model was calculated,  
including all crystal-field and spin-orbit excitations. For each 
composition $x$ the NCA 
%%%%%%%%%%%%%%%%%%%%%%%%%%%%%%%%%%%%%%%%%%%%%%%%%%%%%
\begin{figure}[t]
\begin{center} \scalebox{0.37}[0.37]{\includegraphics[clip]{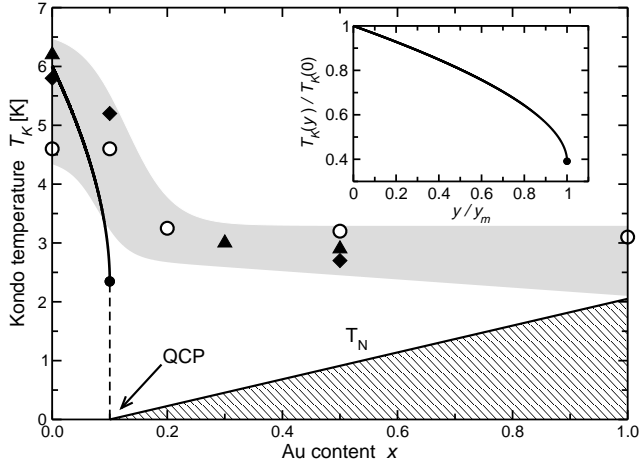}}\end{center}
\vspace*{-0.3cm}
\caption{
Dependence of the Kondo temperature $T_K$ on the Au content $x$, 
as determined by UPS (open circles), specific heat \cite{schlager92} 
(triangles) and neutron scattering 
\cite{stroka93,schroeder00} (diamonds).   
The error bars are approximately the width of the shaded area. 
The N\'eel temperature is labelled by $T_N$. 
The inset and the solid line in the main panel show the 
universal curve $T_K(y)/T_K(0)$ vs. $y/y_{m}$ as given by 
Eq.~(\ref{eq:TK_RG}).
}
\label{fig5}
\end{figure}
%%%%%%%%%%%%%%%%%%%%%%%%%%%%%%%%%%%%%%%%%%%%%%%%%%%%%
spectra are broadened by the experimental resolution and fitted to the 
experimental data, using a single parameter set for all experimental $T$.
Using this parameter set, the NCA spectra were then calculated at low 
temperature, $T\!\approx\!0.1\,T_K$, where $T_K$ was extracted from the 
Kondo-peak half-width at half maximum (HWHM) of the NCA spectra.   
The results are shown in Fig.~\ref{fig5}. 
The finite Kondo scale extracted from the data 
for $x>x_c=0.1$ results from the high-$T$ 
onset of the Kondo resonance seen in Fig.~\ref{fig4}~c)  
which, according to our analysis, is expected not to persist to low $T$. 
Despite an uncertainty in $T_K$, estimated by the width of the shaded area in
Fig,~\ref{fig5}, a sudden decrease of $T_K$ is clearly visible. 
The fact that the sharp $T_K$ drop of the experimental data occurs at 
(or very close to) the quantum critical concentration $x_c$ suggests to 
identify this drop with the theoretically expected signature of the
Kondo breakdown at $y_m$, as illustrated in Fig.~\ref{fig5}. 
This supports that the QPT in CeCu$_{6-x}$Au$_x$ 
follows the LQC scenario driven by intersite magnetic fluctuations, 
as explained in Section \ref{theory}, and as 
was previously inferred from inelastic neutron scattering 
experiments \cite{schroeder00}. However, the nearby orthorhombic-monoclinic 
structural transition occurring at 220~K for $x=0$ and at 70~K for $x=0.1$
\cite{grube99} might also have an effect on the behavior of $T_K$. Therefore,
future work has to substantiate the above LQC conjecture, even though 
across the structural transition the lattice unit cell volume and, hence,
the density of states at the Fermi level tend to {\it increase} smoothly
with increasing $x$, leading to an inrease rather than a drop of $T_K$.

\section{Conclusion}
\label{conclusion}
Our theoretical analysis predicts generally 
that an abrupt step of the Kondo 
screening scale extracted from high-$T$ spectral data
should occur in any HF compound with competing Kondo and RKKY
interactions, as long as the single-ion Kondo screening scale is 
larger than the magnetic ordering temperature. Whether this
distinct feature is located at the quantum critical control parameter
value $x_c$ or inside the magnetically ordered region constitutes a general
high-$T$ criterion to distinguish the LQC and HM scenarios.
Moreover, this criterion allows to predict whether a given system 
should follow the HM or the LQC scenario, once estimates for the
bare single-ion Kondo scale $T_K(0)$ and for  
dimensionless critical coupling strength $y_{SDW}$ for an SDW instability 
in that system are known. This is indicated in Fig.~\ref{fig3} for the
examples of CeCu$_{6-x}$Au$_x$ and CeNi$_{2-x}$Cu$_x$Ge$_2$. A systematical 
analysis of other HF compounds in this respect is in progress. 

We would like to thank F. Assaad, L. Borda, S. Kirchner, A. Rosch and M. Vojta
for fruitful discussions.
This work was supported by DFG through
Re~1469/4-3/4 (M.K., A.N., F.R.), SFB~608 (J.K.) and FOR~960 (H.v.L.).

% \end{multicols}

\end{document}